\documentclass[5p]{elsarticle} % 5 p for preprints
\pdfoutput=1
\usepackage{hyperref,amsmath,amssymb,amsfonts,gensymb,comment,stackengine,graphicx,xcolor,xurl,threeparttable,tabularx,xparse,aas_macros,subcaption,lineno}
\usepackage[nameinlink,capitalise]{cleveref}
\usepackage[normalem]{ulem}
\newcolumntype{Y}{>{\centering\arraybackslash}X}
\usepackage[T1]{fontenc}
\usepackage[utf8]{inputenc}
\usepackage{textcomp,lmodern}
\usepackage[percent]{overpic}

\newcommand{\detname}{\mbox{HP52301-1}}
\newcommand{\ctd}{$\tau_{\textrm{CTD}}$}
\NewDocumentCommand{\quant}{mm}{%
  \ensuremath{#1\,\mathrm{#2}}%
}

\newcommand{\squant}[2]{$\sim$\ensuremath{{#1}}\,\ensuremath{\mathrm{#2}}}
\journal{NIM A}
\crefname{equation}{Eq.}{Eqs.}
\makeatletter
\def\equation@autoref#1{\hyperref[#1]{Eq.~(\ref*{#1})}}
\makeatother

\newcommand{\pitch}{1.162}
\hypersetup{
    colorlinks=true, % Enables coloring of links instead of boxes
    linkcolor=blue,  % Color for internal links (e.g., table of contents, cross-references)
    urlcolor=cyan,   % Color for external URLs
    citecolor=green, % Color for citations
    filecolor=magenta % Color for file links
}
\begin{document}
\begin{frontmatter}

\title{Depth Calibration of Double-sided Strip Germanium Detectors for the \\Compton Spectrometer and Imager Satellite}

\author[SSL]{Field R.\ Rogers\corref{cor1}}
\ead{fieldr@berkeley.edu}
\cortext[cor1]{Corresponding author}
\author[UCSD]{Sean N.\ Pike}
\author[SSL]{Samer Alnussirat}
\author[SSL]{Robin Anthony-Petersen}
\author[UCSD]{Steven E. Boggs}
\author[SSL]{Felix Hagemann}
\author[UCSD]{Sophia E.\ Haight}
\author[SSL]{Alyson Joens}
\author[Goddard]{Carolyn Kierans}
\author[SSL]{Alexander Lowell}
\author[SSL]{Brent Mochizuki}
\author[Goddard]{Albert Y. Shih}
\author[NRL]{Clio Sleator}
\author[SSL]{John A. Tomsick}
\author[SSL]{Andreas Zoglauer}

\address[SSL]{Space Sciences Laboratory, University of California, Berkeley, 7 Gauss Way, Berkeley, CA 94720, USA.}
\address[UCSD]{Department of Astronomy \& Astrophysics, University of California, San Diego, 9500 Gilman Drive, La Jolla, CA, 92093, USA.}
\address[Goddard]{{NASA Goddard Space Flight Center, 8800 Greenbelt Road, Greenbelt, MD, 20771, USA.}}
\address[NRL]{U.S.\ Naval Research Laboratory, 4555 Overlook Ave., SW, Washington, DC, 20375, USA.}

\begin{abstract}
Double-sided strip high-purity germanium detectors with three-dimensional position reconstruction capability have been developed over three decades, with space-based applications in high-energy astrophysics and heliophysics. 
Position resolution in three dimensions is key to reconstruction of Compton scattering events, including for the upcoming Compton Spectrometer and Imager (COSI) satellite mission. 
Two-dimensional position reconstruction is enabled by segmentation of the two detector faces into orthogonal strip contacts, enabling a pixelized analysis.   
The depth of an interaction cannot be measured directly but must be inferred from the charge collection time difference 
between the two faces of the detector. 
Here, we demonstrate for the first time the depth calibration of a detector with the COSI satellite geometry read out using an application specific integrated circuit (ASIC) developed for the COSI mission. 
In this work, we map collection time difference to depth using the Julia-based simulation package \texttt{SolidStateDetectors.jl} and validate it with comparison to the timing distributions observed in data. 
We also use simulations and data to demonstrate the depth resolution on a per-pixel basis, with $>90\%$ of pixels having <\quant{0.9}{mm} (FWHM) resolution at \quant{59.5}{keV} and <\quant{0.6}{mm} (FWHM) resolution at \quant{122.1}{keV}.
\end{abstract}

\end{frontmatter}

%%%%%%%%%%%
% INTRODUCTION %
%%%%%%%%%%%
\section{Introduction}\label{sec:intro}
Technology for double-sided strip high-purity germanium detectors (GeDs) has been developed over multiple decades \cite{Kroeger95,Amman00,Coburn03,Amrose03,Phlips04,Bandstra06,Amman07,Amman18,Sleator19,Amman20}. These sensors combine the excellent energy resolution capability characteristic of germanium with three-dimensional position resolution capability enabled by their geometry. 
Additionally, they are well-suited for $\gamma$-ray detection because the electron density of germanium facilitates a relatively high Compton scattering probability and containment of $\gamma$-rays within a compact instrument. The usefulness of such GeDs for diverse scientific programs and their suitability for high-altitude and space environments has been demonstrated by their successful operation aboard the Gamma-Ray Imager/Polarimeter for Solar flares (GRIPS) \cite{Duncan16}, 
Nuclear Compton Telescope (NCT) \cite{Boggs04,Bandstra11}, and Compton Spectrometer and Imager (COSI) \cite{Kierans16} stratospheric balloons. 

The COSI satellite \cite{Tomsick19,Tomsick22,Tomsick23} is an upcoming compact Compton telescope that will survey the $0.2-5$ MeV sky as a NASA Small Explorer mission. 
Despite a rich scientific potential, MeV-scale astrophysics has thus far been underexplored compared to other energy bands due to the experimental challenges of operating in the Compton scattering regime -- including the need for a high-altitude or space-based mission, high instrumental and astrophysical backgrounds, and low interaction cross sections. COSI's 16 GeDs enable the energy and position resolution necessary for reconstruction of Compton scattering events. 
They thus facilitate sensitivity for MeV astrophysics, especially narrow lines, despite the challenges. The energy and position resolution of the GeDs directly impact COSI's angular resolution, critical for imaging capability and background suppression \cite{Lowell16}.

\begin{figure}[tb]
\centering
\includegraphics[width=0.48\textwidth,trim = 80 0 150 0,clip]{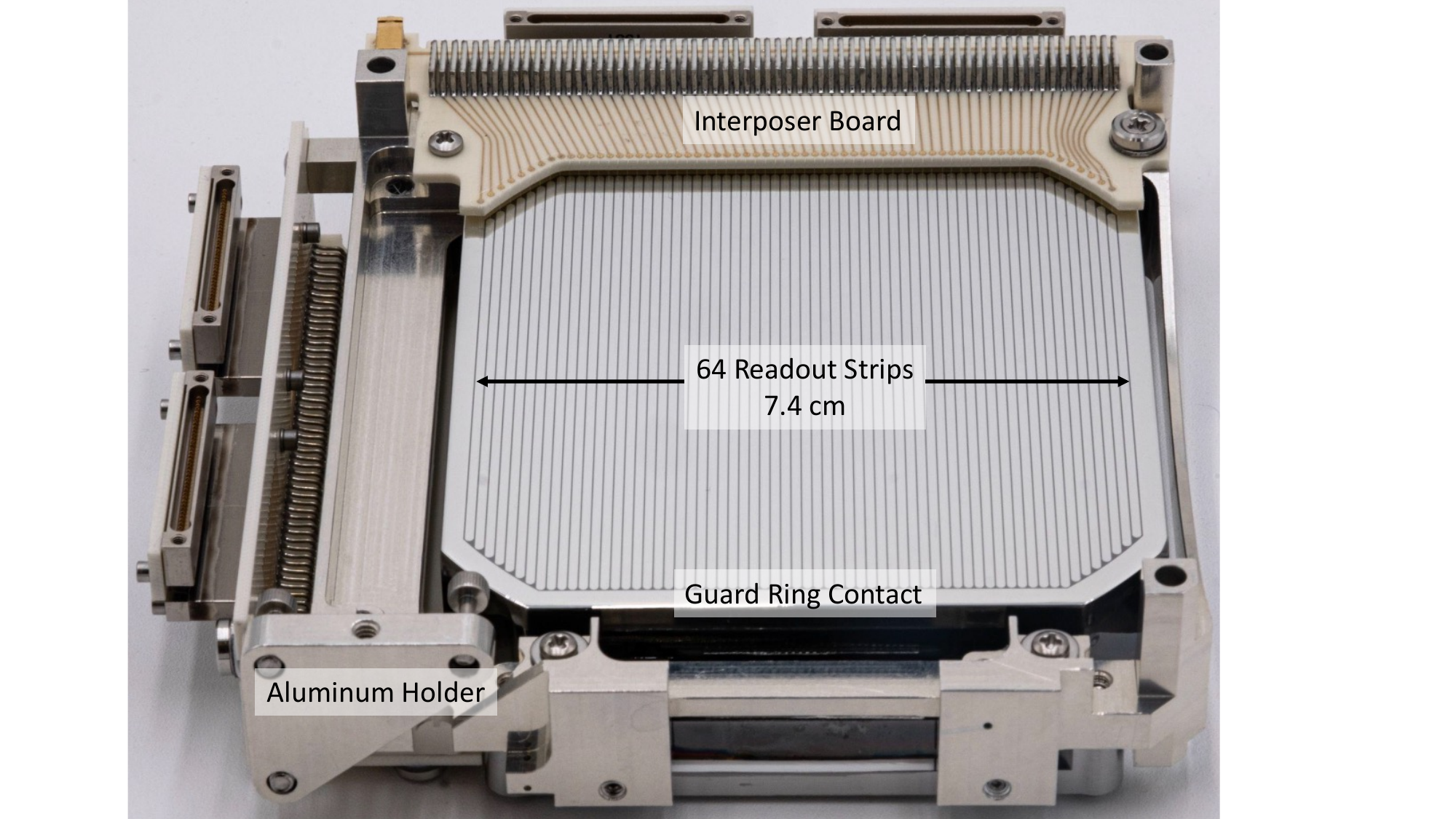}
\caption{\label{fig:ged}
A 64-strip GeD in an engineering-model aluminum holder. {The interposer board} and connectors at the top of the image carry signal from the 64 HV strip contacts (visible and oriented vertically in this image), while the connectors on the left carry signal from the 64 LV strip contacts (hidden, and oriented horizontally). The guard ring contact encloses the 64 strip contacts on each face. 
}
\end{figure}

Each COSI satellite GeD (pictured in \autoref{fig:ged}) features a \squant{15}{mm}-thick planar geometry. Each face is segmented into 64 strip contacts (\quant{\pitch}{mm} strip pitch), with the strips on the low voltage (LV; cathode) face orthogonal to those on the positive high voltage (HV; anode) face. 
Defining the intersection of a unique HV and LV strip pair as a \quant{\pitch\times\pitch}{mm^2} pixel for analysis purposes enables direct 2D position reconstruction as well as a pixelized calibration. 

At the energies considered in this work, an incident $\gamma$-ray typically interacts either via Compton scattering or photoabsorption. 
In the case of Compton scattering, it may scatter one or more times prior to either photoabsorption in or escape from the GeD. 
At each interaction site, $\sim$340 electron-hole pairs are generated per keV, resulting in a cloud of free charge carriers. 
The electron cloud and hole cloud propagate toward opposite faces of GeD at typical velocities of a few \quant{10^6}{cm/s}, depending on the temperature, electric field strength, and material properties. As they propagate, the charge clouds also expand due to Coulomb repulsion and diffusion.

The motion of the charges induces a signal on the contacts. A weighting potential \cite{He21} is defined to describe the signal induced on a particular contact by charge movement at some position. 
The COSI GeDs are subject to the small-pixel effect \cite{Barrett95}, where the signals on small contacts are dominated by charge cloud motion close to the contacts. Thus, the signal on any HV (LV) contact is predominantly induced by the motion of electrons (holes) as they approach that contact. 
Effects including electronic and environmental noise, charge trapping in the bulk, charge sharing between strips, signal loss to the gaps between strips, capacitive cross talk, and transient signals all contribute to the GeD response \cite{Liu09,Beechert22,Pike23,Boggs23fitting,Boggs23mod,Boggs23num,Pike25}. 

The depth ($z$) of an interaction cannot be measured directly. Instead, $z$ must be inferred from the charge collection time difference (\ctd), defined as the time elapsed between the signal on the HV contact and the  signal on the LV contact.  
The \ctd-$z$ relation depends on factors including operating bias voltage, impurity concentration, detector thickness, and temperature-dependent charge carrier mobility in the material. 
The small pixel effect means that the weighting potential depends strongly on $z$ near the faces of the GeD, so the relation is nonlinear\footnote{Additionally, for events close to the HV (LV) face, the holes (electrons) also contribute to the signal, resulting in a steep slope near the faces.}. 
Thus, the \ctd-$z$ mapping must be calculated or simulated for each GeD at its operating temperature and bias voltage. 

For calibration purposes, it is possible to either a) place a flood source at a known position, model the distribution of interaction depths relative to the source position, and compare the data to the model or b) use a collimated source or fan beam to illuminate some of the edge pixels at a known depth. 
This work utilizes a flood source at a known position. 
The distribution of energy depositions as a function of $z$ can be calculated analytically or simulated using a particle simulation toolkit such as \texttt{Geant4} \cite{geant4}, where the latter method can more easily incorporate the effects of scattering in passive materials. 

Depth calibration for similar GeDs has previously been demonstrated \cite{Amman00,Coburn03,Amrose03,Bandstra06,Liu09,Lowell16}. 
Here, we present the first depth calibration of a GeD with the COSI satellite geometry with \quant{\pitch}{mm} strip pitch, and the first  using the COSI satellite-model readout electronics. 
We also demonstrate the use of the Julia-based \cite{julia} \texttt{SolidStateDetectors.jl}\footnote{\href{https://github.com/JuliaPhysics/SolidStateDetectors.jl}{https://github.com/JuliaPhysics/SolidStateDetectors.jl}} \cite{SolidStateDetectors} framework to simulate the timing response of these double-sided strip GeDs. 
\autoref{sec:data} presents the experimental setup including modeling of $z$ distributions and the preparation of the calibration data. 
\autoref{sec:sims} details the simulations of electric field and weighting potentials, charge cloud transport within the GeD, the readout electronics, leading to a \ctd-$z$ mapping and a model \ctd\ distribution. 
In \autoref{sec:cal}, we fit the data with the model to obtain the \ctd-$z$ mapping and the depth resolution for a single pixel, and in \autoref{sec:full}, we discuss the results from fitting data from all of the pixels individually. 

% -------------- -------------- -------------- 
\section{Data Collection}\label{sec:data}

\subsection{Detector, Readout, and Operation}\label{sec:det}
This work demonstrates the depth calibration of a single COSI satellite engineering-model GeD, \detname, with its readout electronics. \detname\ was fabricated at Lawrence Berkeley National Laboratory using amorphous Si and Ge contact technology \cite{Amman07,Amman18,Amman20}. 
Its physical characteristics are summarized in \autoref{tab:ged}. 
As illustrated in \autoref{fig:ged}, each COSI satellite GeD is a \squant{15}{mm}-thick crystal cut from a \quant{10}{cm}-diameter boule. 
The instrumented area is the intersection of an \squant{8}{cm} square with this circular geometry. 
The \quant{7.44}{cm}-wide fiducial area is segmented into 64 readout strips on the LV face and 64 orthogonal strips on the HV face. 
A guard ring contact encloses the strips on each side of the detector, improving the noise performance by mitigating surface leakage current to the strips. The geometry is distinct from previous GeD models in strip number and pitch.

The operational setup, in which a single GeD is held within a custom liquid-nitrogen-cooled vacuum cryostat, is based on that in Ref.~\cite{Amman20}. The temperature of the GeD is \squant{83}{K}, extrapolated based on a sensor located on a metal bracket holding the detector holder assembly within the cryostat. 
\detname~was operated under an external \quant{1000}{V} bias, compared to the depletion voltage V$_\textrm{D}\sim175$\,V.

Each GeD is read out by six 32-channel custom application-specific integrated circuit (ASIC) chips \cite{Wulf20}, which provide energy and timing information for each strip within the power, volume, and data bandwidth constraints of a space mission. All six ASICs are controlled by a single field-programmable gate array (FPGA) and respect its clock. The strips are read out by four ASICs, two AC-coupled to the HV strips, and two DC-coupled to the LV strips. 
The LV and HV guard ring contacts are  read out by one channel each of the final two ASICs. 
In each ASIC channel, the signal is first processed by a preamplifier circuit with an \quant{8.33}{MHz} low-pass order-2 Butterworth filter. Then, the preamplifier output is processed by two different shaping circuits. 
The `slow shaper', an order-5 Gaussian filter with peaking time set to \quant{2}{}\textmu s, is used for the energy scale, where the analog-digital converter (ADC) records the peak amplitude of the slow shaper output. 
Meanwhile, the `fast shaper,' an order-3 Gaussian filter with  peaking time set at \quant{60}{ns}, is used to calculate the timing of charges arriving at the strip. When the fast shaper peaks, the ASIC begins to ramp an analog voltage for that readout channel. 
The time-amplitude conversion (TAC) is the digitized value of the voltage at the end of the readout window for an event. 
A `slow threshold' (typically \squant{10}{keV}) and a `fast threshold' (typically \squant{35}{keV}) are independently set for the slow and fast shaper on each channel. 

All six ASICs are triggered when any readout channel exceeds its slow threshold.  
ADC and TAC values are then recorded for each ASIC channel exceeding its slow threshold within a fixed window and for all adjacent strips. 
For each channel, flags indicate if the slow and fast thresholds were exceeded. 
If the fast threshold is not exceeded for a channel, then the fast shaper does not trigger a voltage ramp, and instead the ASIC begins to ramp the voltage at the peak of the slow shaper, resulting in a smaller TAC value. 
Each energy deposition results in charge on both the HV and LV faces, such that the resulting data typically consists of at least one channel over threshold on either face of the GeD.

\begin{table}[tbp]
\caption{\label{tab:ged} The physical characteristics and operation conditions of GeD \detname.}
\centering
\vspace{1mm}
\renewcommand{\arraystretch}{1.1} % more vertical spacing
\begin{tabularx}
{0.95\columnwidth}{ll}
\hline
Number of strips per side & {64} \\ 
Strip contact width &  \quant{1.0166}{mm} \\
Strip gap width & \quant{0.1452}{mm} \\
Guard ring width & \quant{2.5}{mm}*\\
Crystal lateral dimension & \quant{79.5}{mm} \\
Crystal thickness & \quant{15.2}{mm} \\
Contact thickness & \quant{0.5}{}\textmu\textrm{m} \\
Impurity type & p-type \\
Impurity concentration & \quant{1.3\cdot10^9}{cm^{-3}}\\
Estimated depletion voltage, V$_{\textrm{D}}$& \quant{175}{V}\\ 
Operating bias voltage & \quant{1000}{V}\\
Operating temperature & \quant{83}{K}\\
Crystal Orientation in \^z & $\langle$001$\rangle$\\
\hline
\end{tabularx}
{\\ *\small \it The guard ring contact width varies around the GeD perimeter as determined by the precise positioning of the photolithography mask used to apply the strips.}\\
\end{table}

\subsection{Datasets and Event Selection}\label{sec:reduction}

This section outlines the depth calibration datasets, energy and timing calibration, and the event selection leading to the generation of \ctd\ distributions from data.

The data used in the depth calibration were obtained using $^{241}$Am and $^{57}$Co sources placed outside of the cryostat, at a distance roughly \quant{20}{cm} from either the HV or the LV face of the GeD and roughly centered with respect to the face. Photons interacting in matter are characterized by an energy-dependent attenuation coefficient $\lambda$, such that the intensity $I(r)$ of a photon beam that has traveled distance $r$ in a material is given by 
\begin{equation}\label{eq:attenuation}
    I(r) = I(0)\cdot e^{-r/\lambda}.
\end{equation}
The number of photoabsorption events in the range ($r$,$ r + \delta r$) is proportional to $I(r)$. For each source, \autoref{tab:sources} lists the attenuation coefficients for $\gamma$-rays at the line energies of interest in this calibration.  

For the pixelated GeD geometry with a source placed at a distance from the center of the HV or LV face, the effective 
$I(z)$ in a pixel is modified relative to \cref{eq:attenuation}, due to a) the angular dependence of the coherent scattering cross section, b) the angle between $\hat{r}$ and $\hat{z}$ for off-center pixels, c) the $z$-dependence of the solid angle subtended by a pixel. 
\texttt{Geant4} simulations demonstrated that for the source positioned at \squant{20}{cm}, $I(z)$ in a given pixel still follows an approximately exponential distribution in depth, with an effective attenuation depth in a central pixel of $\lambda_z^c\sim 0.98 \lambda$ ($0.90\lambda$)
at \quant{59.5}{keV} (\quant{122.1}{keV}). Depending on their exact position, the effective attenuation depth $\lambda_z$ for non-central pixels differs from $\lambda_z$ at the percent level. 

\begin{table}[tbp]
\caption{\label{tab:sources}The radioisotopes and selected line energies used in the depth calibration. The mass attenuation coefficients ($\lambda/\rho$) are from the NIST XCOM database \cite{xcom}, neglecting coherent scattering. The attenuation coefficients in mm ($\lambda$) are calculated using the Ge density of \quant{\rho =  5.323}{g/cm}$^{3}$. $\lambda_z^c$ is the effective attenuation depth in a central pixel based on \texttt{Geant4} simulations.}
\centering
\vspace{1mm}
\renewcommand{\arraystretch}{1.1} % more vertical spacing
\begin{tabularx}{0.95\columnwidth}{YYYYY}%c}%c
\hline
Isotope& %Activity & 
Energy & $\lambda/\rho$ & $\lambda$ & $\lambda_{z}^c$\\
& [keV] & [cm$^2$g$^{-1}$] & [mm] & [mm]\\%& [mm]\\
\hline
$^{241}$Am &59.5&1.945 & 0.97 & 0.95 \\%
$^{57}$Co & 122.1 & 0.327  &  5.75 & 5.20\\
\hline
\end{tabularx}
\end{table}

The energy and TAC calibrations were generated and applied on a per-strip basis. %, resulting in energy values in keV and time values in ns. 
The energy calibration was generated for each strip of \detname ~using \texttt{Melinator} photopeak-fitting software\footnote{\texttt{Melinator} is packaged with the Medium-Energy Gamma-ray Astronomy library (\texttt{MEGAlib}), available online at \href{https://github.com/zoglauer/megalib}{https://github.com/zoglauer/megalib}.} \cite{Zoglauer06} and dedicated datasets recorded with $^{241}$Am, $^{57}$Co, and $^{137}$Cs calibration sources following the procedure detailed in previous work \cite{Sleator19,Beechert22}. 
Meanwhile, the TAC calibration was calculated for each strip by injecting \quant{50}{mV} pulses (corresponding to \squant{50}{keV} signals) at known times after a trigger, to obtain a linear conversion from raw clock values to nanoseconds. 

Low-level cuts were imposed on the strip-hits to ensure data quality. 
Strip-hits that did not exceed the slow thresholds were not considered in this analysis. Requirements were then placed on the timing for each strip-hit within an event. 
A software coincidence window of \quant{600}{ns}, much longer than the observed $\sim$\quant{200}{ns} maximum charge drift time for \detname, was imposed based on TAC values. 
Additionally, to remove chance coincident events, strip-hits were required to have TAC values corresponding to the window when the ASICs were accepting events rather than reading out data. 

Event selection criteria were applied to obtain a clean sample of single-site events with a well-defined energy: 
\begin{enumerate}
\item{\bf Energy threshold:} An \quant{8}{keV} (\quant{17}{keV}) software energy threshold was imposed on the strip-hits in the $^{241}$Am ($^{57}$Co) data to mitigate impacts of strip-to-strip threshold variation in the subsequent selection\footnote{The software threshold is $\sim$14\% of the line energy in both cases, which allows some charge to be shared onto adjacent strips. Because there is more charge sharing at higher energies and the degree of charge sharing depends on the interaction depth, this choice also mitigates depth-dependent selection effects when requiring exactly one strip-hit per face.}.
\item{\bf Single-pixel events:} Events were selected with exactly two strip-hits above the energy threshold, one strip-hit on either face. 
\item{\bf Line energy:} To select a clean sample of photopeak events, events were selected for which one of the two strip-hits fell within $\pm$\quant{3.5}{keV} of the line energy (i.e., within \squant{3}{\sigma} of the line centroid considering the energy resolution on a typical strip). The other strip was required to be less than the line energy + \quant{3.5}{keV}. This relatively permissive selection allows events with sub-threshold charge sharing or trapping on one side of the detector to be included in the analysis, mitigating systematic depth-dependent effects in the event selection.
\item{ \bf Fast timing:} Events were rejected if either of the constituent strip-hits did not exceed its fast threshold and thus lacked a valid fast-shaper TAC value. 
\end{enumerate}

\begin{figure}[tb]\centering
\begin{overpic}[width=0.5\textwidth,trim=16 55 13 14,clip]{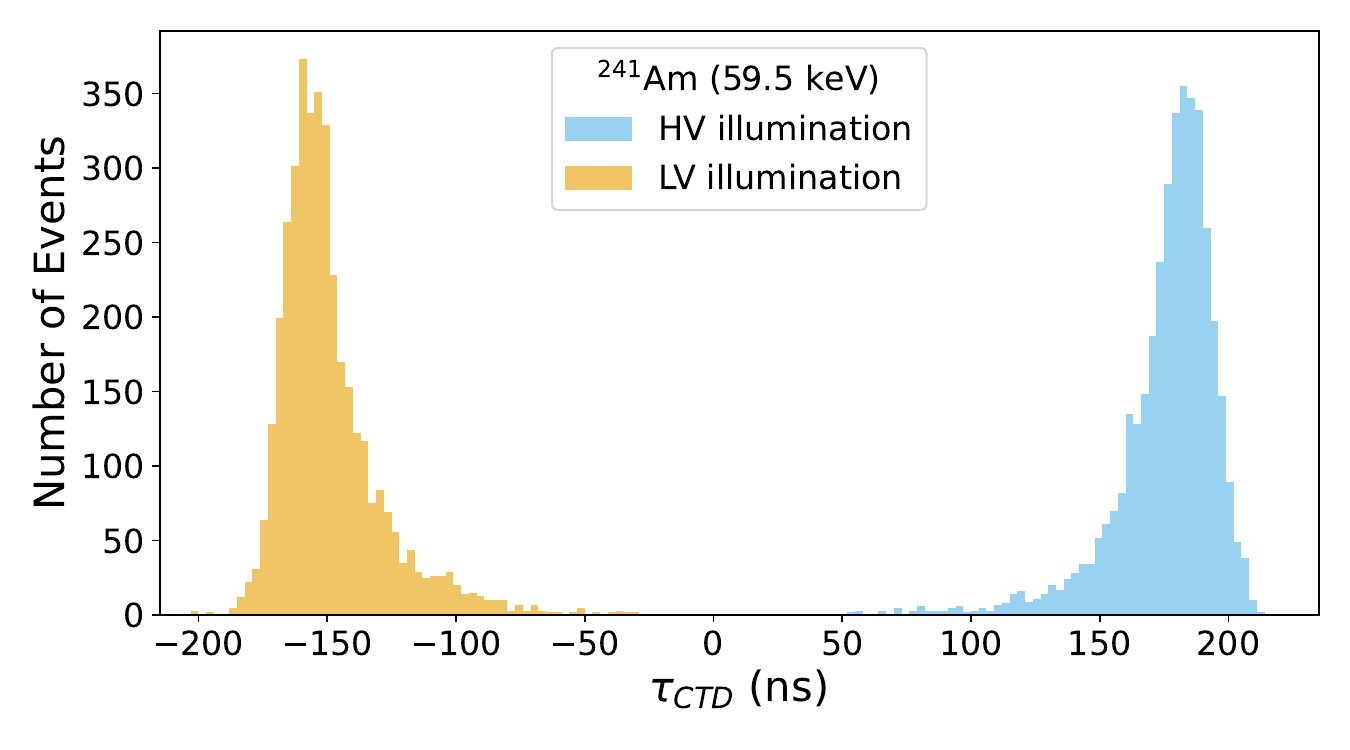}
\put(12,40){(a)} % example label
\end{overpic}
\begin{overpic}[width=0.5\textwidth,trim=16 19 13 14,clip]{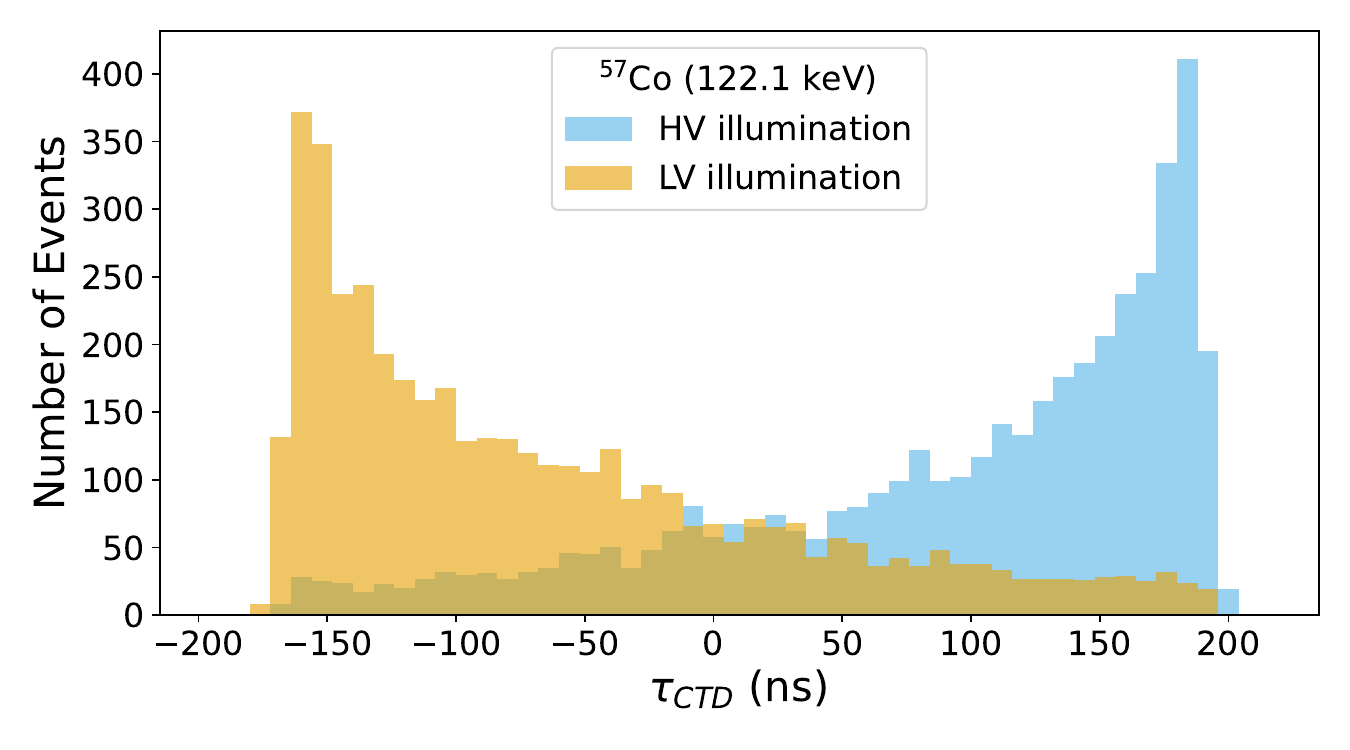}
\put(12,46){(b)} % example label
\end{overpic}
\caption{\label{fig:data}
\ctd\ distributions for an example pixel in \detname\ are shown for events in the \quant{59.5}{keV} line of $^{241}$Am (a; \quant{3}{ns} bin width) and the \quant{122.1}{keV} line of $^{57}$Co (b; \quant{8}{ns} bin width). For each source, data was collected with the source illuminating the GeD from the LV face (orange, peaked at  \ctd \quant{\sim-160}{ns}) and HV face (blue, peaked a \ctd\ \quant{\sim180}{ns}).
}\end{figure}

For each event passing the selection criteria, \ctd\ was calculated as the difference between the calibrated TAC values on the HV strip and the LV strip. 
\autoref{fig:data} presents the resulting \ctd\ distributions for an example pixel for both sources and both source positions. 
For the line energies in \autoref{tab:sources}, $\lambda_z$ is smaller than the GeD width. 
Thus, the \ctd\ distributions peak near the edges, corresponding to energy depositions near the illuminated face. 
At \quant{59.5}{keV}, the events are entirely concentrated near the illuminated face, while at \quant{122.1}{keV}, the entire detector width is illuminated for both source positions. 

\section{Simulation of Collection Time Difference}\label{sec:sims}

\subsection{Collection Time Difference to Depth Mapping}\label{sec:julia}

This section describes the solid state detector and pulse shaping simulations that were used to create a mapping between \ctd\ and interaction depth $z$. 

A GeD with the geometry and operating parameters summarized in \autoref{tab:ged} and material characteristics given in \autoref{tab:sim} was simulated using the Julia language-based \texttt{SolidStateDetectors.jl} simulations package \cite{SolidStateDetectors}. The coordinates were defined with the GeD centered on ($x,y,z$) = (0,0,0) and aligned with the edge of LV (HV) strip 63 at \quant{x = +37.1}{mm} (\quant{y = +37.1}{mm}) and the HV face at \quant{z = +7.6}{mm}. 
The electric potential and electric field were numerically computed throughout the sensitive volume, and the weighting potential was computed for each contact. 

\begin{table}[tbp]
\caption{\label{tab:sim} Parameters in the \texttt{SolidStateDetectors.jl} simulation.}
%\centering
\vspace{1mm}
\renewcommand{\arraystretch}{1.1} % more vertical spacing
\begin{tabularx}{0.95\columnwidth}{ll}
\hline
%\hline
Ionization energy& \quant{2.95}{eV}\\
Fano factor & 0.129 \\ 
Density & \quant{5.323}{g\,cm^{-3}}\\
Dielectric constant & \quant{16.0}{{\epsilon}_0} \\
Electron diffusion coefficient & \quant{239}{cm^2 s^{-1}}\\
Hole diffusion coefficient & \quant{279}{cm^2 s^{-1}}\\
Initial charge cloud radius & \quant{20}{}\textmu m\\
Longitundinal effective electron mass & \quant{1.64}{m_e}\\
Transverse effective electron mass & \quant{0.0819}{m_e}\\
\hline
\end{tabularx}
\end{table}

\begin{figure}[tb]\centering
\includegraphics[width=0.48\textwidth,trim = 9 5 7 0, clip]{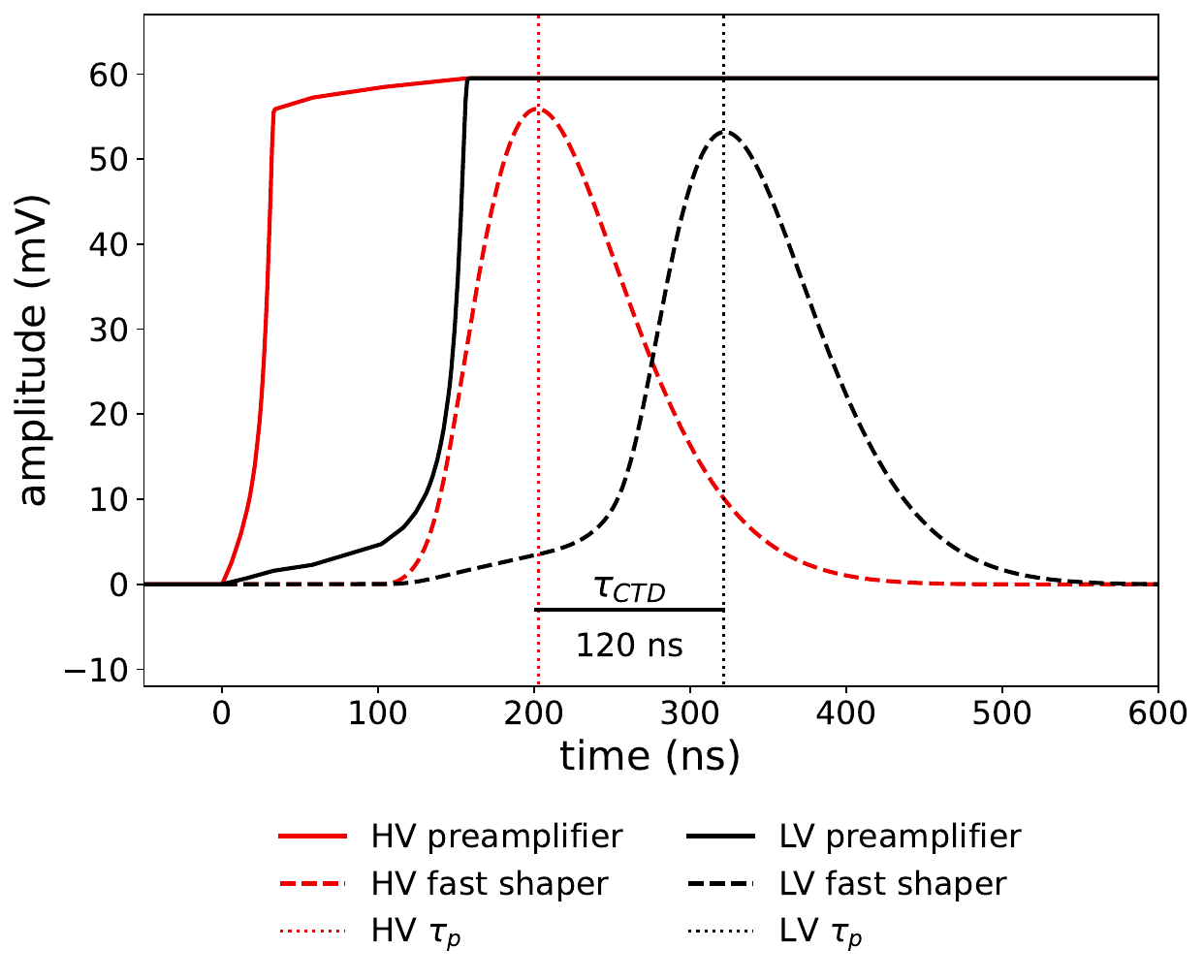}
\caption{\label{fig:waveforms}
Simulated waveforms on the HV (red) and LV (black) contacts resulting from a charge cloud generated at \quant{z = +4.6}{mm} in the detector and centered laterally on the example pixel. For both contacts, simulated pulses are illustrated after processing through the preamplifier (solid) and fast shaper (dashed). The dotted vertical lines indicate $\tau_p$, the start time for the TAC clocks at the peak of the shaped signal. \ctd\ is calculated according to \cref{eq:CTD} at \quant{120}{ns}. 
}\end{figure}

\autoref{fig:waveforms} illustrates the simulated GeD and ASIC response to a single energy deposition. 
Electron and hole clouds corresponding to a \quant{59.5}{keV} energy deposition were generated with $(x,y,z)$ = (\quant{+0.58}{mm}, \quant{+0.58}{mm}, \quant{+4.60}{mm}), centered within the area of a sample pixel and \quant{3.0}{mm} from the HV contact. 
The propagation of the electrons and holes was simulated in the \texttt{SolidStateDetectors.jl} framework, including diffusion and self-repulsion effects. 
A charge carrier drift velocity model with a powerlaw temperature dependence was used, with electron and hole mobility defined following   Ref.~\cite{Bruyneel06}. 
The charge induced on the contacts was recorded as a function of time, with \quant{1}{ns} sampling. 
To produce the semi-Gaussian pulses in \autoref{fig:waveforms},  waveforms from the \texttt{SolidStateDetectors.jl} simulation were processed through a 5-channel SPICE-based\footnote{\href{https://ptolemy.berkeley.edu/projects/embedded/pubs/downloads/spice/index.htm}{https://ptolemy.berkeley.edu/projects/embedded/pubs/\\downloads/spice/index.htm}} \cite{Nagel73} simulation of the ASIC electronics, and the simulated fast-shaper output was plotted for the channels of interest. 

In \autoref{fig:waveforms}, the signal on the HV strip rises \squant{120}{ns} before the signal on the LV strip. This is because the charge cloud was generated \quant{3.0}{mm} (\quant{12.2}{mm}) from the HV (LV) face, so in this example the holes traverse a longer distance to the LV contact than the electrons to the HV contact. 
The vertical lines indicate the time $\tau_p$ of the peak of the fast shaper. 
In the ASIC, the TAC clock starts at $\tau_p$, and the TAC value for that channel is its clock time at the end of the event. \ctd\  is defined as:
\begin{equation}\label{eq:CTD}
   \tau_{\textrm{CTD}} := \textrm{TAC}_{HV}-\textrm{TAC}_{LV} = \tau_{p,LV}-\tau_{p,HV}.
\end{equation}
For the example in \autoref{fig:waveforms}, the signal on the HV strip rises before the LV strip,  $\textrm{TAC}_{HV} > \textrm{TAC}_{LV}$, and \ctd$>0$.

\begin{figure}[tb]\centering
\includegraphics[width=0.48\textwidth,trim = 25 0 60 45,clip]{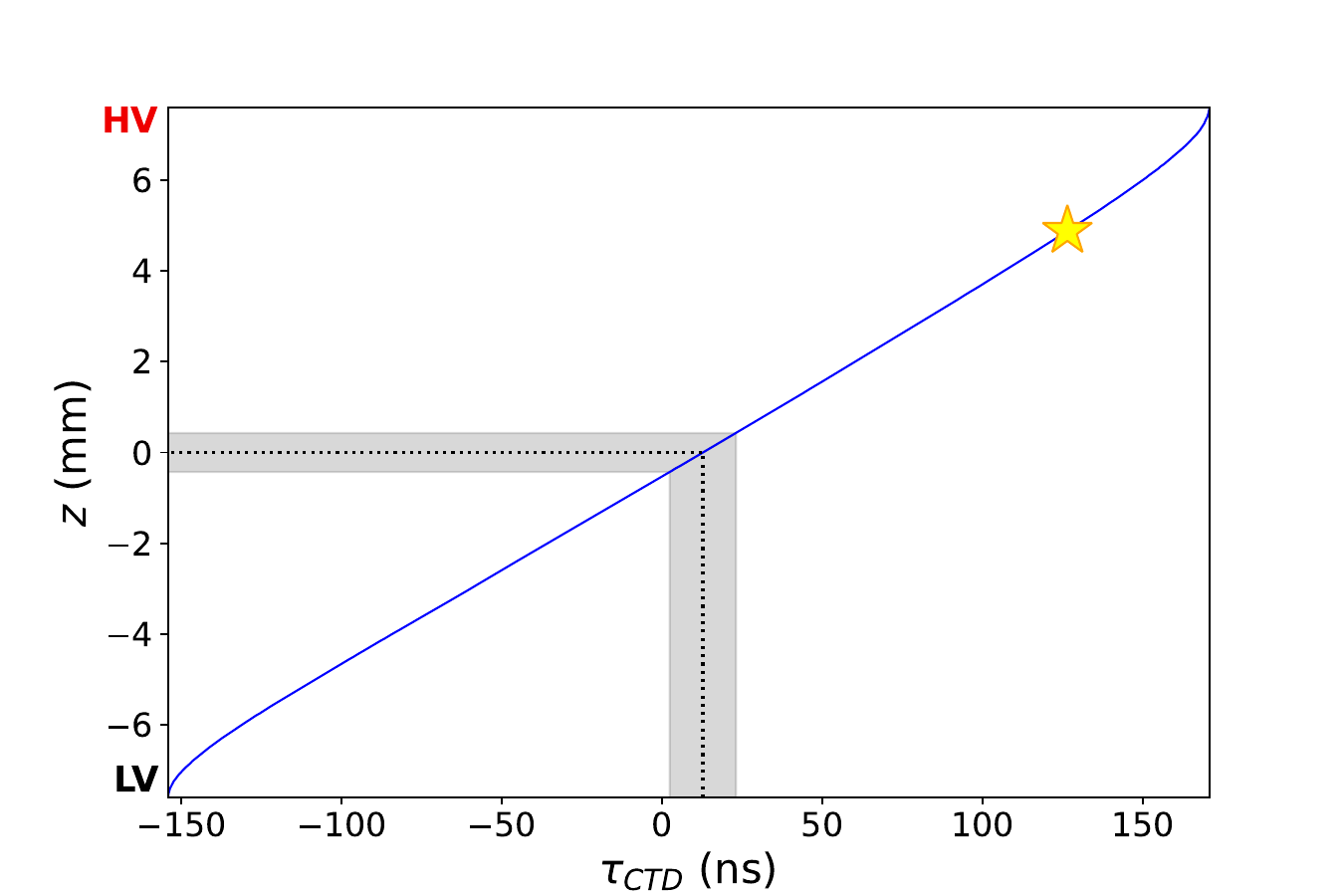}
\caption{\label{fig:ctd_vs_depth}
The simulated relation between charge collection time difference (\ctd; defined in \cref{eq:CTD}) and interaction depth ($z$) is illustrated in blue. The HV face is at \quant{z = +7.6}{mm}. The waveform illustrated in \autoref{fig:waveforms} constitutes a single point (indicated by the star) on this curve. The depth resolution ($\Delta z$) is calculated at depth $z$ by mapping $z$ to \ctd\ (black dotted line), adding the \ctd\ resolution (gray band) and mapping the band back onto depth. 
}\end{figure}

\autoref{fig:ctd_vs_depth} illustrates the relationship between \ctd\ and $z$. 
Charge clouds were injected at \quant{0.05}{mm} increments across the \quant{15.2}{mm} thick GeD. 
For each charge injection, the waveforms were simulated and processed as in \autoref{fig:waveforms}. 
The resulting curve shows an approximately linear relation across much of the crystal width, with a steepening (increased $\delta z$/$\delta $\ctd) near the faces of the GeD due to the sharp weighting field arising from the small-pixel effect as well as charge induced by electrons (holes) on the hole-collecting (electron-collecting) contacts.

\subsection{Simulated \ctd\ Distributions}\label{sec:cosima}

In this section, the exponential distribution of interaction depths discussed in \autoref{sec:reduction}, the \ctd-$z$ map calculated in \autoref{sec:julia}, and a Gaussian timing resolution are combined to simulate a \ctd\ distribution for a given energy and source position.

The ideal \ctd\ distribution was modeled for the $\lambda_z^c$ values in \autoref{tab:sources} and the two source positions by defining 1000 bins spanning the simulated \ctd\ range, mapping the respective exponential depth distribution into those bins, convolving with a Gaussian timing resolution $\sigma_{\textrm{CTD}}$, and linearly interpolating between the points using the \texttt{interp1d} functionality from \texttt{SciPy} \cite{scipy}. 
The result is a \ctd\ probability density function (PDF) that depends on the effective attenuation coefficient $\lambda_z$ and the timing resolution $\sigma_{\textrm{CTD}}$. 

\begin{figure*}[tb]\centering
\includegraphics[width=0.48\textwidth,trim = 10 0 10 10,clip]{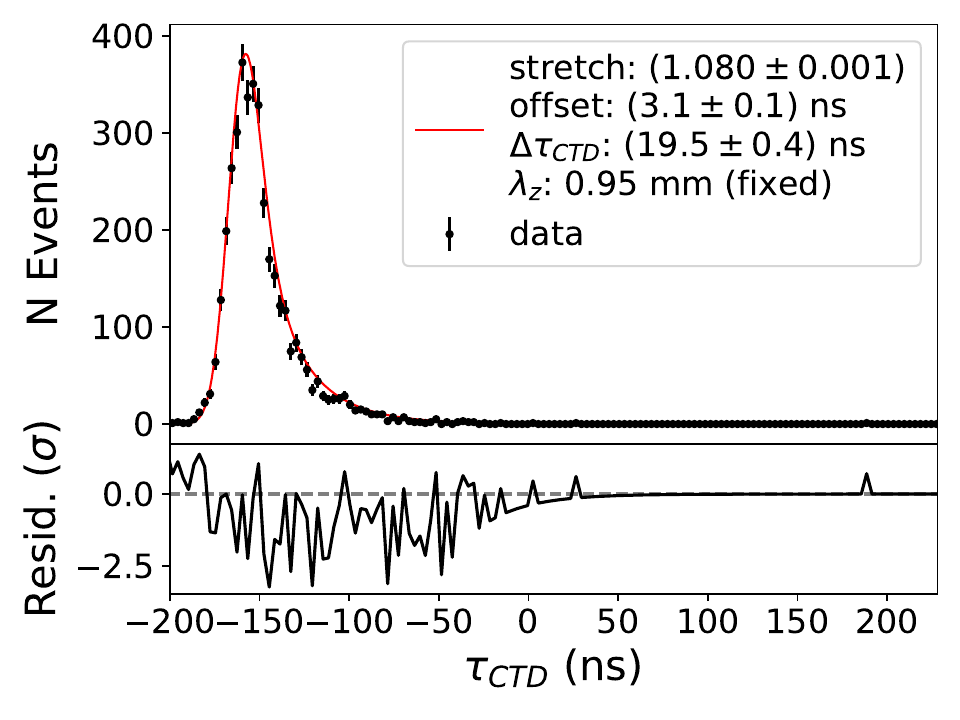}
\includegraphics[width=0.48\textwidth,trim = 10 0 10 10,clip]{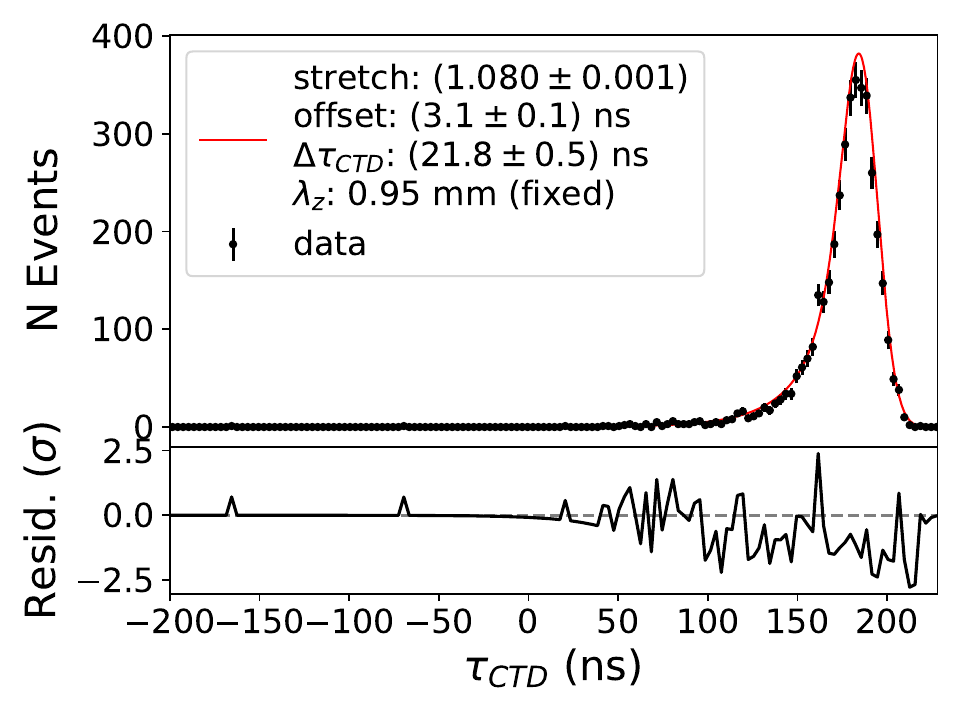}\\
\includegraphics[width=0.48\textwidth,trim = 10 0 10 10,clip]{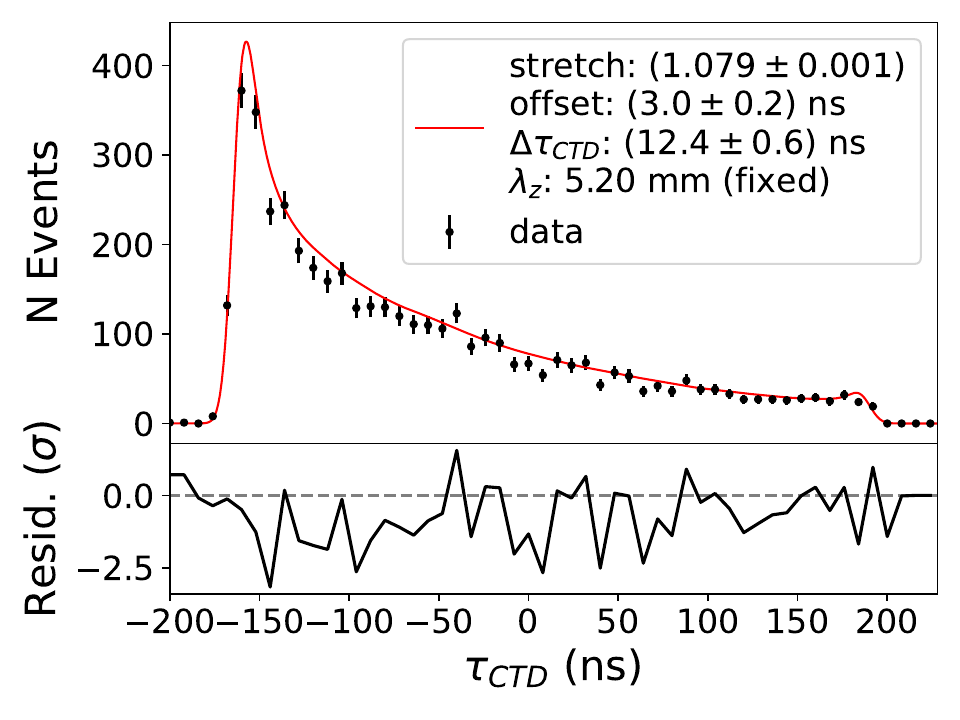}
\includegraphics[width=0.48\textwidth,trim = 10 0 10 10,clip]{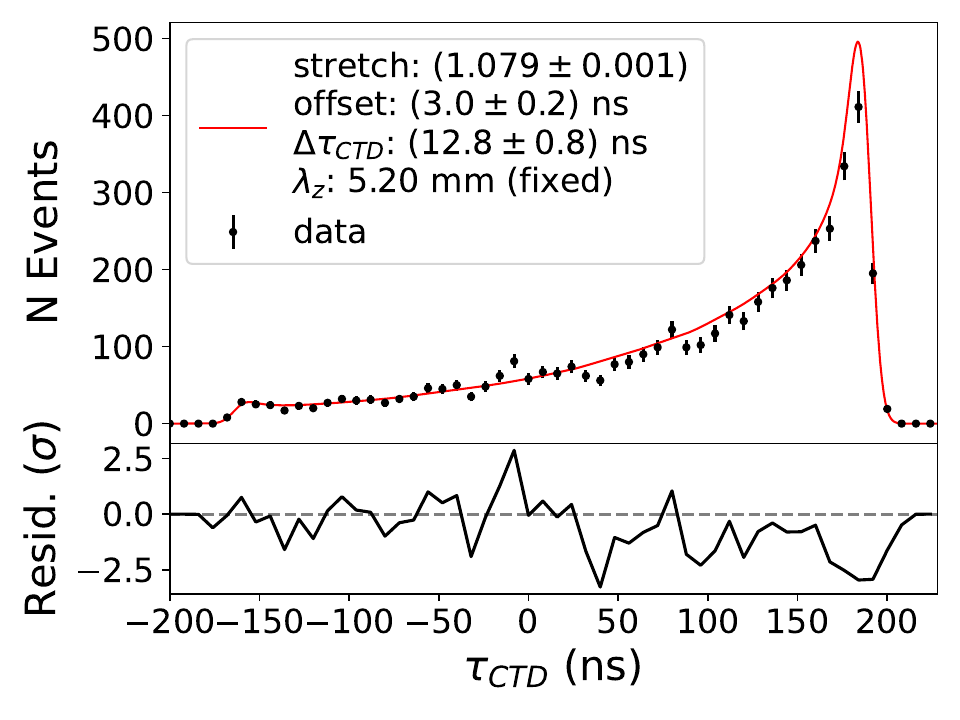}\\
\caption{\label{fig:fits}
Fitted \ctd\ distributions for a single pixel illuminated with $^{241}$Am (upper) and $^{57}$Co (lower) from the LV (left) and HV (right) faces. 
The $^{57}$Co and $^{241}$Am fits were performed independently, but resulted in consistent fitted values for the stretch  and offset. For each source, distributions from the two source positions were fitted jointly, with only $\Delta$\ctd\ allowed to vary between the LV- and HV-illuminated data. Residuals are reported based on the most probable number of counts in a bin according to the model. }\end{figure*}

\section{Per-pixel Depth Calibration}\label{sec:cal}

In this section, the \ctd\ distributions for each individual pixel from \autoref{sec:reduction} were fitted with a PDF generated as in \autoref{sec:cosima} to validate the simulated \ctd-$z$ map and to extract the timing resolution and thus the uncertainty on a depth reconstruction. 

First, to facilitate fitting of data from the nearly 4000 pixels in the GeD, cumulative density functions (CDFs) were generated spanning $2\,\textrm{ns} < \sigma_{\textrm{CTD}} < 30\,\textrm{ns}$ with \quant{0.05}{ns} step size for each $\lambda_z$ and illuminated face. The CDF as a continuous function of $\sigma_{\textrm{CTD}}$ was defined by interpolating between the pre-calculated curves.  

For each pixel and for each source with its characteristic line energy, a binned negative log-likelihood cost function was defined simultaneously relating the HV- and LV-illuminated data to the respective model.
For each fit, an offset and a stretch parameter were defined such that 
\begin{equation}\label{eq:stretchoff}
    \tau_{\textrm{CTD,data}} = (\tau_{\textrm{CTD,sim}} + \textrm{offset}) \cdot \textrm{stretch}.
\end{equation}
Varying stretch and offset in the fit allows for variations in material properties or crystal width across the lateral extent of the GeD \cite{Lowell16}, as well as errors on the charge mobility parameters in the charge-drift model. 
While stretch and offset were common between the HV- and LV-illuminated data, $\sigma_{\textrm{CTD}}$ varied independently for the two distributions, for a total of four free parameters. 
To avoid biasing the fit, bins with fewer than 5 counts were masked prior to minimization. 
The cost function was reduced using the \texttt{migrad} fitting routine from \texttt{iMinuit} \cite{iminuit2.30.1}, and errors were evaluated using \texttt{minos}.

\begin{figure}[tb]\centering
\includegraphics[width=0.48\textwidth,trim = 9 0 63 51,clip]{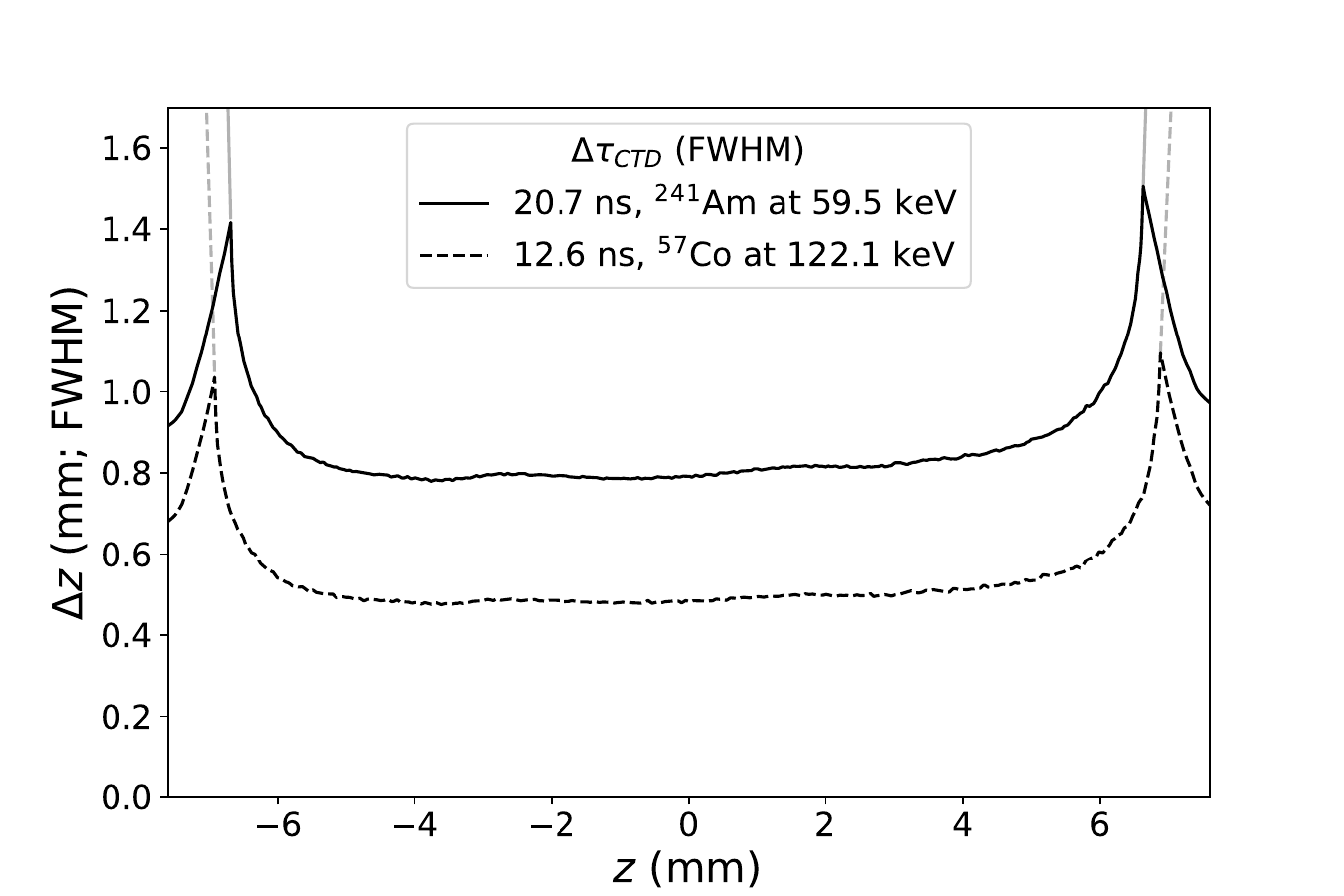}
\caption{\label{fig:depth_stat}
Depth resolution ($\Delta$z; FWHM) as a function of interaction depth ($z$), for the mean $\Delta$\ctd\ obtained for the pixel in \autoref{fig:fits} at \quant{59.5}{keV} (solid) and \quant{122.1}{keV} (dashed). For this pixel, $\Delta z = $ \quant{0.79}{mm} (\quant{0.48}{mm}) at \quant{z = 0}{mm} for $\Delta$\ctd\ = \quant{20.7}{ns} (\quant{12.6}{ns}). For 70\% of the GeD thickness, \quant{-5.8}{mm} $ < z < $ \quant{4.9}{mm}, $\Delta z$ is within 10\% of the \quant{z = 0}{mm} level.
}\end{figure}

\autoref{fig:fits} illustrates the \ctd\ data and best fit simulated curve for both sources, for the example pixel from \autoref{fig:data}. 
Overall, the ability of the simulated \ctd\ distributions to fit the data validates the use of the simulated \ctd-$z$ mapping for depth calibration in this GeD. 
The best fit stretch and offset are consistent between the fits at the two energies. The 8\% stretch factor indicates that the real charge drift times are 8\% longer than predicted in simulation, compared to an $\sim10\%$ error on the simulated charge drift mobility values reported in several works \cite{SolidStateDetectors,Caughey67,Mihailescu00}. 
The \squant{3}{ns} offset could indicate mismodeling of the relative drift velocity for electrons vs holes but is also less than the reported \quant{5.5}{ns}  uncertainty in the TAC calibration offset for this pixel and thus can also be entirely explained by calibration error. 
The fitted stretch and offset are further discussed in \autoref{sec:full} in the context of the full GeD. 

The reported $\Delta$\ctd\ is the full-width at half maximum timing resolution,  calculated as $2.355\cdot\sigma_{\textrm{CTD}}\cdot\textrm{stretch}$. 
The best-fit $\Delta\tau_{CTD}$ values are inconsistent between the HV- and LV-illuminated data at \quant{59.5}{keV}. This could indicate that assumptions of the model --- such as the assumption of constant $\Delta\tau_{CTD}$ over the depth of the GeD --- may be incorrect, or that the fit is being distorted by variable correlations. Because the \quant{59.5}{keV} photons travel only short distances into the GeD, the \ctd\ distributions at \quant{59.5}{keV} pick up effects within \squant{1-2}{mm} of the faces, while the \quant{122.1}{keV} distributions both probe the entire thickness. 
Comparison of the result at \quant{59.5}{keV} and \quant{122.1}{keV} indicates an energy-dependence of the timing resolution, which is finer at higher energies.

\autoref{fig:depth_stat} illustrates the depth resolution ($\Delta z$; FWHM) as a function of interaction depth for the example pixel fitted in \autoref{fig:fits}. 
As illustrated in \autoref{fig:ctd_vs_depth}, $\Delta z(z)$ was calculated by mapping $z$ to \ctd, finding the \ctd\ range corresponding to $\Delta$\ctd, and mapping the range of \ctd\ values back into $z$, accounting for stretch and offset in comparison with the simulated map. 
This mapping was performed using the mean of the HV- and LV- best-fit $\Delta$\ctd\ values at \quant{59.5}{keV} and \quant{122.1}{keV}. 
The finer $\Delta$\ctd\ at the higher-energy line corresponds to a finer $\Delta z$. 
Over most of the detector's thickness, $\Delta z$ is nearly independent of $z$. However, $\Delta z$ increases near the contacts, where small changes in \ctd\ result in large changes in reconstructed $z$. The gray curves indicate the value of $\Delta z$ if $z$ is allowed to be reconstructed at any value of $z$. The sharp peaks occur as an artifact when energy depositions are constrained to reconstruction within the thickness of the GeD.

\section{Full Detector Depth Calibration}\label{sec:full}

The fitting procedure in \autoref{sec:cal} was performed for every pixel. This section discusses results for the GeD overall. Of the 3984 pixels, the 64 pixels of LV strip 14 were excluded due to anomalous behavior of the ASIC channel. Additionally, one pixel was excluded from the $^{241}$Am data and ten pixels were excluded from the $^{57}$Co data because the fits failed to converge.

\begin{figure}[tb]\centering
\includegraphics[width=0.48\textwidth,trim = 9 0 63 48,clip]{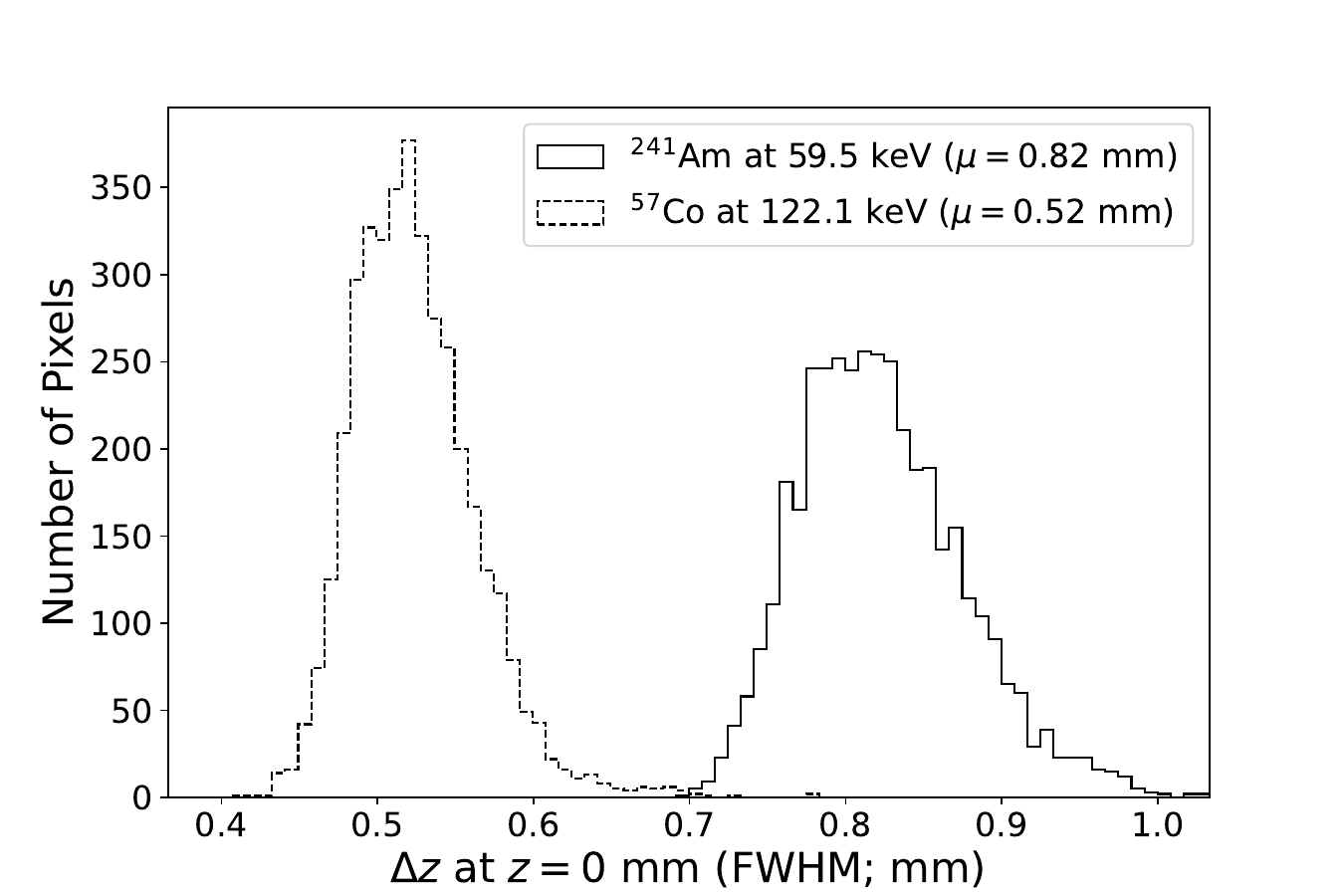}
\caption{\label{fig:res_stat}
$\Delta z$ (FWHM; \quant{z = 0}{mm}) for all pixels with converging fits. The calculation was based on the mean of the HV- and LV-fitted $\Delta$\ctd\ at each energy.
}\end{figure}

\autoref{fig:res_stat} illustrates $\Delta z$ (FWHM) at \quant{$z = 0$}{mm} for the \detname\ pixels for the $^{241}$Am line at \quant{59.5}{keV} and the $^{57}$Co line at \quant{122.1}{keV}. As demonstrated in \autoref{fig:depth_stat}, $\Delta z$ at \quant{z=0}{mm} is representative of $\Delta z$ over most of the width. 
At \quant{59.5}{keV} (\quant{122.1}{keV}), the average pixel has \quant{\Delta z =0.82}{mm} (\quant{\Delta z=0.52}{mm}). \autoref{fig:res_stat} indicates a significant energy dependence in $\Delta z$. Finer $\Delta$\ctd, and thus finer $\Delta z$, is anticipated at higher energies, where the signal-to-noise ratio on the fast shaper is more favorable. Full modeling of the energy dependence is deferred to a future study. 

At a given energy, $\Delta$\ctd\ for the HV- and LV-illuminated \ctd\ distributions are systematically different and linearly correlated. The trend is most stark for $^{241}$Am, where the regression $\Delta$\ctd$_{,LV}$ = 0.9$\cdot\Delta$\ctd\quant{_{,HV} + 0.6}{ns} gives $\sigma = $ \quant{0.3}{ns} on the residuals. 
The effect is less significant for $^{57}$Co, possibly because the entire GeD thickness is illuminated with both source positions at \quant{122.1}{keV} or because of the lower quality of the fits overall. 
Though the difference between $\Delta$\ctd\ on the HV vs the LV face is a $\sim$10\% effect, small compared to the overall $\Delta$\ctd, this result indicates that the assumption of constant $\Delta$\ctd\ for a pixel at a given energy does not necessarily hold, and motivates a deeper future investigation into the position dependence of $\Delta$\ctd.

\begin{figure*}[tb]\centering
 % \begin{subfigure}[t]{0.48\textwidth}
    \centering
\includegraphics[width=0.48\textwidth,trim = 30 32 0 85,clip]{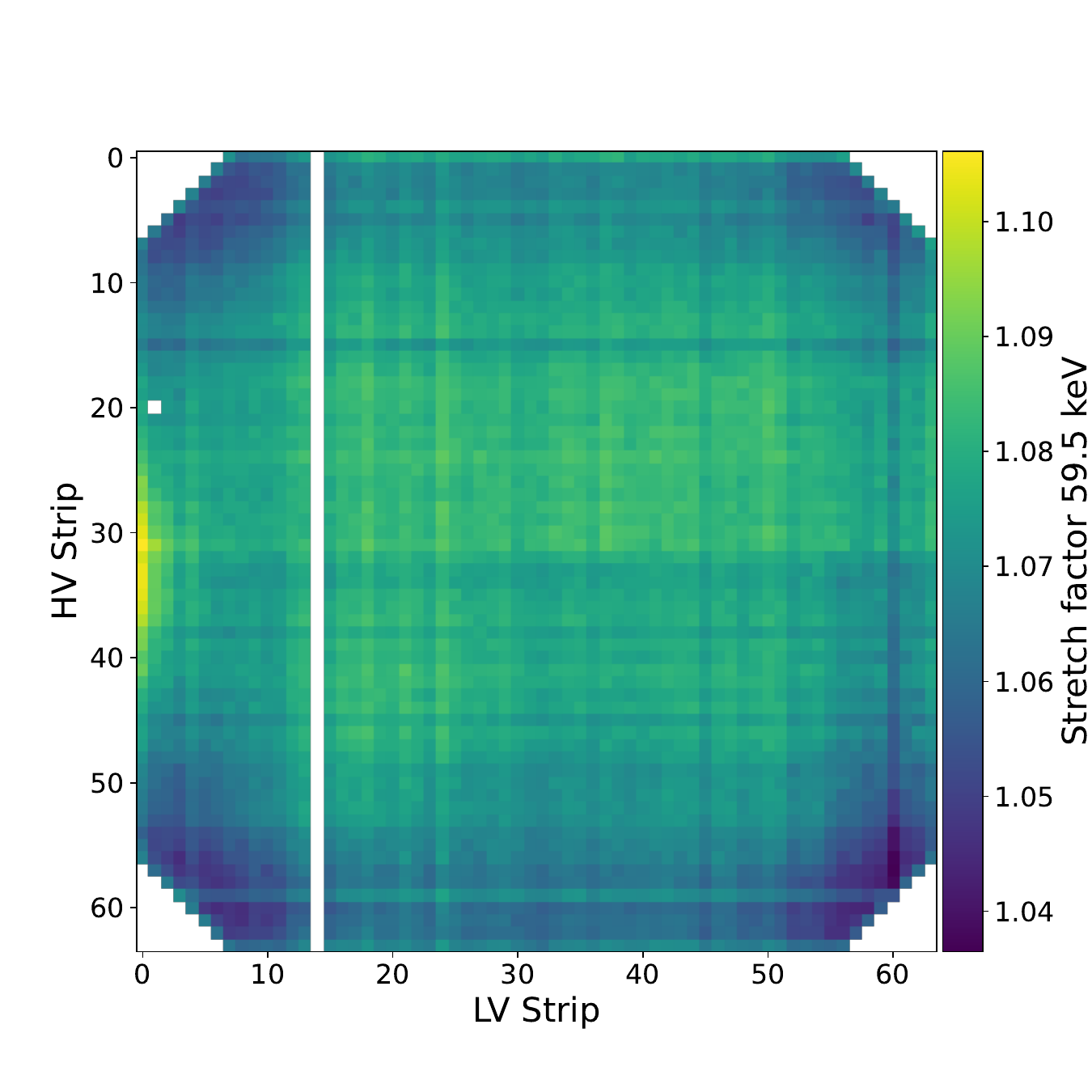}  \includegraphics[width=0.48\textwidth,trim = 27 32 3 85,clip]{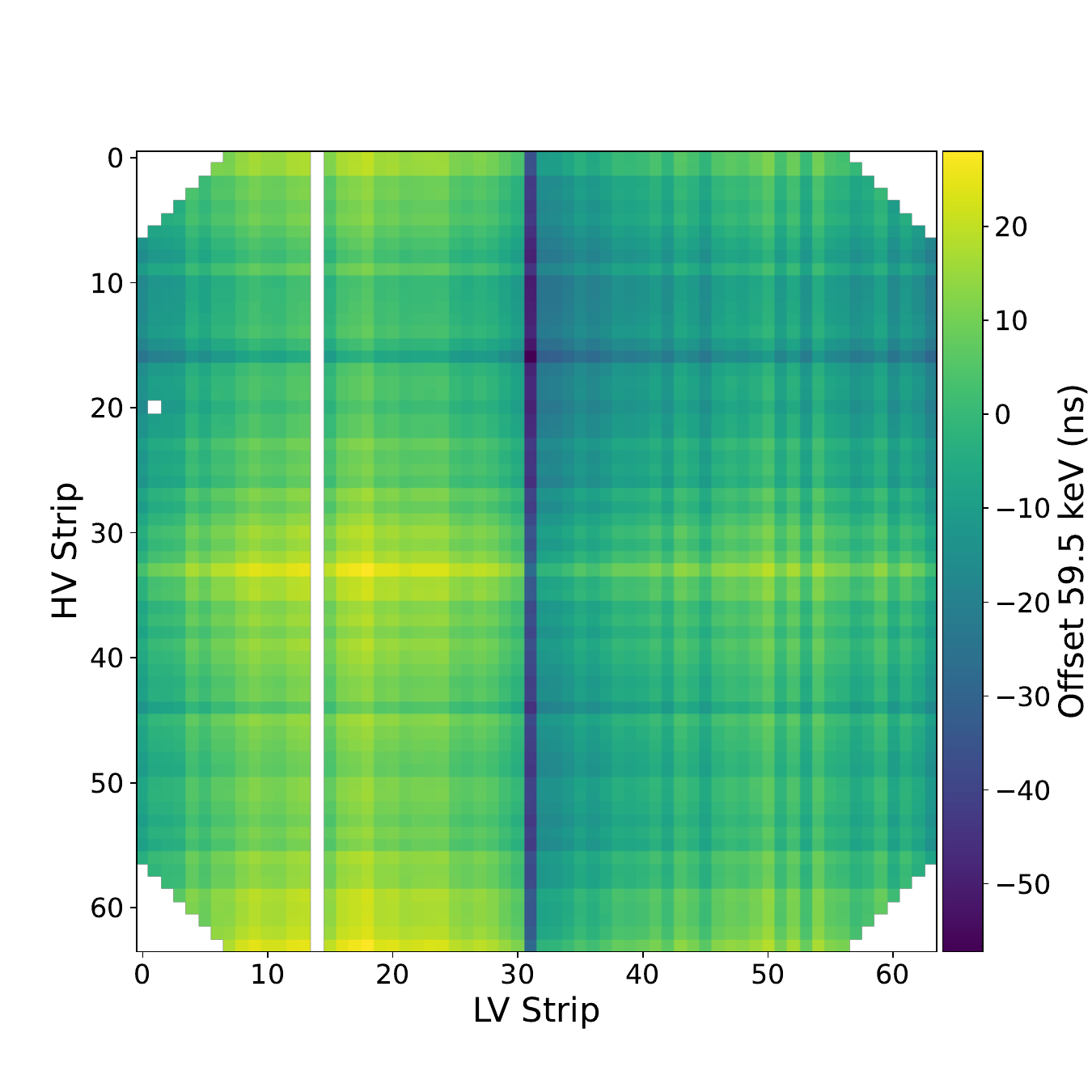}  \includegraphics[width=0.48\textwidth,trim = 30 32 0 85,clip]{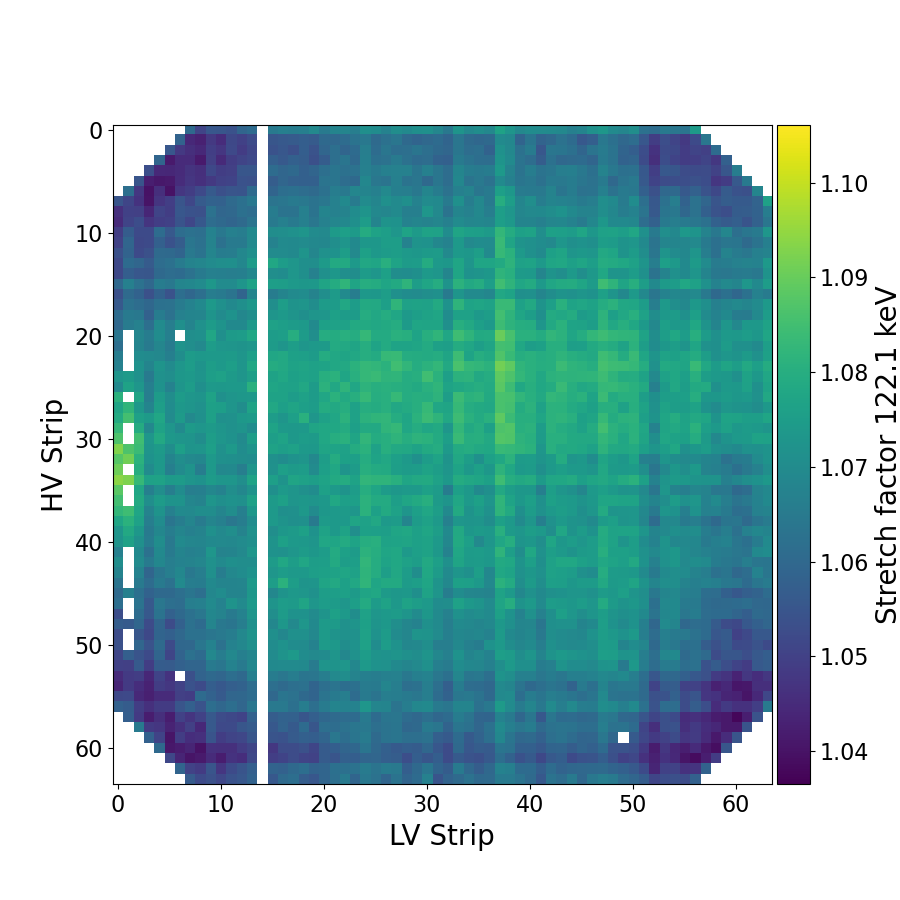} %\caption{}
%    \label{fig:headstretch}
%    \end{subfigure}
%  \begin{subfigure}[t]{0.48\textwidth}
%    \centering
\includegraphics[width=0.48\textwidth,trim = 27 32 3 85,clip]{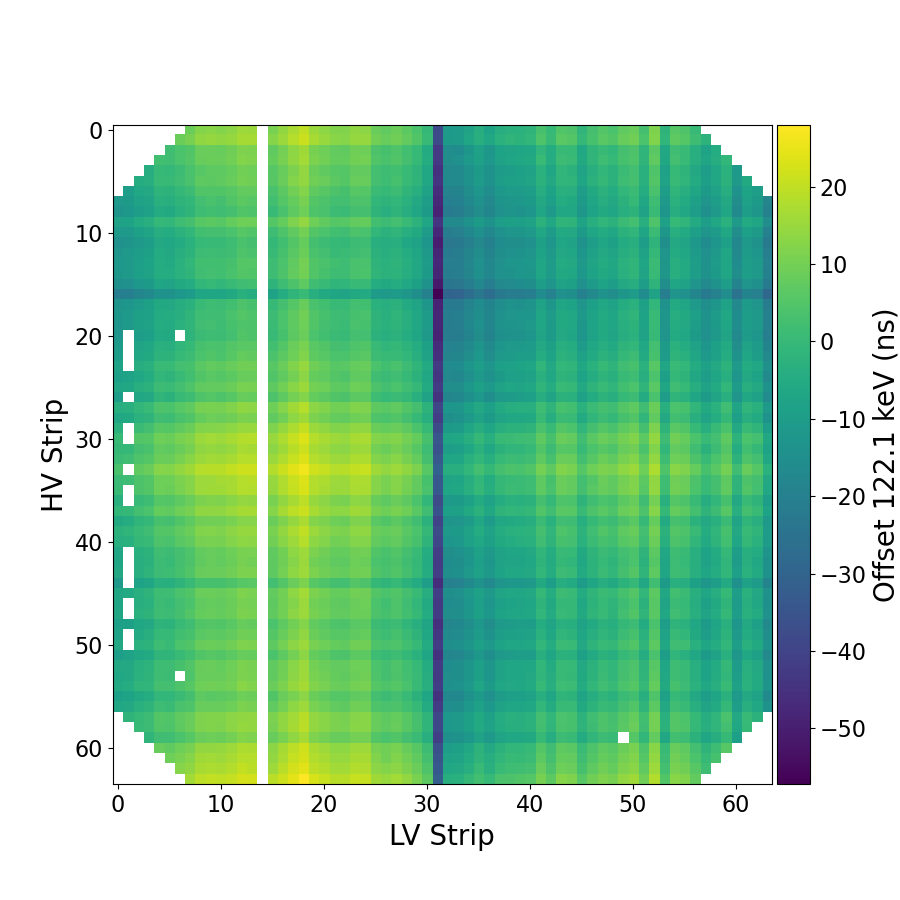} %\caption{}
%    \label{fig:heatoffset}
%    \end{subfigure}
\caption{\label{fig:heat}
Pixel map of the best-fit stretch (left) and offset (right) from \cref{eq:stretchoff} in $^{241}$Am (upper) and $^{57}$Co (lower). As illustrated in \autoref{fig:ged}, the 7 strips nearest each edge  have fewer than 64 pixels per strip in order to fit into the crystal geometry, resulting in the missing corner pixels. At both energies, the stretch factor varies by <10\% across the GeD. A radially-dependent pattern is apparent, where the stretch factor is higher (longer drift times) toward the center of the detector and smaller at larger radii, except that the edge pixels feature anomalously high stretch. By contrast, the offset values appear primarily determined by the HV and LV strips rather than a spatial pattern over the detector surface, indicating that the variation primarily originates in the TAC calibration, which is performed on a per-strip basis.}
\end{figure*}

\autoref{fig:heat} illustrates the fitted stretch and offset from $^{241}$Am and $^{57}$Co, mapped across the lateral extent of the GeD. 
A slight energy dependence in the stretch was observed, as the mean percent stretch over all of the pixels was 1.073 (1.068) at \quant{59.5}{keV} (\quant{122.1}{keV}) with pixel-to-pixel variation $\sigma_{\textrm{stretch}} =  0.009$ at both energies. 
If this effect is real, its impact on the reconstructed $z$ is a \quant{<1}{ns} effect at \ctd\ \quant{= 180}{ns}, much less than $\Delta$\ctd, but it still motivates further study at higher energies. 
At either energy, the result is somewhat smaller than the expected $\sim$10\% error in the charge mobility model \cite{SolidStateDetectors,Caughey67,Mihailescu00}. Error in the measured GeD thickness, in the extrapolated temperature, or in the TAC calibration could contribute to this deviation, though these contributions are expected to be at the sub-percent level. A population of pixels in LV strip 0 with anomalously long drift times (larger stretch) may indicate a local defect in the crystal structure or an impact of the guard ring electrode or nearby Al holder on the electric fields. 
Overall, the charge drift time is greater for the central pixels and decreases at larger radii, except for the very edge pixels. This suggests lateral inhomogeneities in physical characteristics such as the impurity concentration or crystal thickness, or edge effects on the electric field.
The longer drift times in the edge pixels are more likely related to non-uniform electric fields near edges of the crystal or near the larger guard ring contacts. 

For both energies, the mean offset is \quant{0}{ns}, with an \quant{11}{ns} standard deviation over the pixel population. {The offsets are also very strongly correlated between the two fits for a given pixel.} This indicates that the drift velocity model is correctly reproducing the relative difference in the electron and hole drift times, but some other factor is responsible for the pixel-to-pixel variation. The reported uncertainty in the TAC calibration per strip indicates an expected \quant{8}{ns} standard deviation in the \ctd\ offsets based only on TAC calibration uncertainties. Examination of the strip-by-strip trends in \autoref{fig:heat} supports the hypothesis that most of the fitted offset for a typical pixel can be described as the combination of a calibration `offset' from each of its constituent strips.

\section{Conclusion}\label{sec:conclusion}
The ability to precisely reconstruct the 3D position of energy depositions in detectors is critical for Compton event reconstruction and thus for Compton imaging. Development of this capability in double-sided strip GeDs has enabled MeV imaging capability with a compact Compton telescope design. In the COSI GeDs, position in $x$ and $y$ is primarily determined by associating energy depositions with an orthogonal pair of \quant{\pitch}{mm}-wide strips. The depth $z$ of interaction is extrapolated from the relative arrival time of charge at either face.

This work departs from previous depth calibration of similar GeDs both in terms of the hardware and in terms of the methods. 
The GeD in this work differs from those in previous work in the number of strips and in the strip pitch. Additionally, this is the first demonstration of the depth calibration using the COSI satellite-model ASIC readout. 
This work also represents the first depth calibration for a double-sided cross-strip GeD with \ctd-$z$ mapping based on the \texttt{SolidStateDetectors.jl} simulation package, which was validated for timing simulations in the COSI GeDs based on compatibility with measured \ctd\ distributions. 

The depth resolution has been reported for two low-energy $\gamma$-ray lines, \quant{59.5}{keV} and \quant{122.1}{keV}. This is in contrast to previous work, which used a continuum spectrum from the scattering of \quant{661.7}{keV} photons from $^{137}$Cs   \cite{Lowell16}. The use of two distinct lines  highlights the energy dependence of the timing resolution, and by extension the depth resolution, in this energy range. 

This study motivates several deeper investigations into the timing characteristics of the COSI GeDs, including characterization of:
\begin{itemize}
\item the energy- and depth-dependence of the timing resolution; 
\item the impacts of dead layers, impurity gradients in $z$, and other effects that could affect electric fields near the contacts;  
\item the impacts of lateral impurity gradients and thickness-nonuniformity on the drift velocity characteristics across the GeD; and 
\item edge effects, guard ring effects, and other factors that could impact electric fields near the edge of the GeD. 
\end{itemize}
Additionally, this work selected events for which the line energy was primarily contained within a single strip. The depth calibration method for the case of charge sharing between adjacent strips requires treatment of the systematic timing effects of moving charges on neighboring strips. This study has been deferred to a future work. Finally, validation of the method using a few pixels illuminated by a fan beam is in progress and will also be reported in a future publication.

This work has demonstrated a method for calibrating the third positional dimension in the COSI satellite GeDs. Timing data from the ASIC has been mapped onto the depth of interaction, with a unique mapping and fitted depth resolution for each pixel. 
90\% of the pixels had depth resolution <\quant{0.9}{mm} at \quant{59.5}{keV} and <\quant{0.6}{mm} at \quant{122.1}{keV} at the center of the GeD's thickness. 
The calibration results from this GeD motivate detailed characterization of the timing characteristics of the COSI GeDs and ASICs.

%acknowledgments
\section*{Acknowledgment}
The Compton Spectrometer and Imager is a NASA Explorer project led by the University of California, Berkeley with funding from NASA under contract 80GSFC21C0059.

We gratefully acknowledge the effort by the developers of the \texttt{SolidStateDetectors.jl} simulation framework, which has made this work possible. 
We also thank the members of the COSI collaboration's instrument and data pipeline teams for many enlightening discussions. F.R.\ thanks Martha E.\ Field for proofreading the manuscript. 

Software: \texttt{SolidStateDetectors.jl} \cite{SolidStateDetectors}, \texttt{MEGAlib} \cite{Zoglauer06}, \texttt{SPICE} \cite{Nagel73}, \texttt{Geant4} \cite{geant4},
\texttt{iMinuit} \cite{iminuit2.30.1}, 
\texttt{Matplotlib} \cite{matplotlib}, 
\texttt{SciPy} \cite{scipy}, 
\texttt{NumPy} \cite{numpy}, \texttt{Pandas} \cite{pandas,pandasSW}.

\bibliographystyle{naturemag_noURL}
\bibliography{cosi}

\end{document}